# Polymorphic Worms Collection in Cloud Computing

Ashraf A. Shahin[1, 2]

[1]Department of Computer Science, Al Imam Mohammad Ibn Saud Islamic University (IMSIU), Kingdom of Saudi Arabia

[2]Department of Computer and Information Sciences, Cairo University, Egypt

ashraf_shahen@ccis.imamu.edu.sa

*Abstract— In the past few years, computer worms are seen as one of significant challenges of cloud computing. Worms are rapidly changing and getting more sophisticated to evade detection. One major issue to defend against computer worms is collecting worms' payloads to generate their signature and study their behavior. To collect worms' payloads, we identified challenges for detecting and collecting worms' payloads and proposed high-interactive honeypot to collect payloads of zero-day polymorphic worms in homogeneous and heterogeneous cloud computing platforms. Virtual machine (VM) memory and VM disk image are inspected from outside using open-source forensics tools and VMWare Virtual Disk Development Kit. Our experiments show that the proposed approach overcomes the identified challenges.*

## I. INTRODUCTION

Rapid growth of the demand for computational outsourcing has led to the creation of large-scale cloud computing data centers. Organizations outsource their computational needs to cloud data centers instead of incurring high upfront costs in purchasing IT infrastructure, and dealing with maintenance and upgrades of both software and hardware [2]. Cloud computing data centers leverage virtualization technology to allow creation of multiple Virtual Machine (VM) instances on a single physical server and allow creation of multiple virtual networks (VN) on a single physical network [1].

Cloud computing networks contain thousands of VMs. Such networks are susceptible to attack from attackers and malware programs, so that the cloud providers have to take pre-emptive steps to ensure security in their cloud networks. Computer worms are the most danger malware computer programs that harm cloud networks [4, 15]. Worms can be used to perform many malicious activities (e.g., steal information, launch-flooding attacks against servers, etc.) [14]. Worms' writers try to make their worms undetectable as long as possible to infect many more hosts. In another side, cloud providers try to minimize worms' life and spreading in their cloud networks.

Using virtualization software (e.g., VMWare, Xen, and KVM), all components in the real physical networks can be virtualized, and spreading of worms can be minimized by isolating the traffic of each virtual network from the traffic of other networks. The traffic is isolated even if their VMs are on the same physical server and their virtual interfaces transmit traffic across the same underlying NIC [6]. Any VM outside of the virtual network cannot sniff or inject traffic into the virtual network [6]. However, VMs are still susceptible to attack from worms running in VMs in the same virtual network.

Detecting and collecting worms' payloads are very significant processes in defending against unknown worms. Collected payloads can be used to generate worms' signatures and to study their malicious activities. In this paper, we identified a set of challenges for detecting and collecting worms' payloads. Additionally, we proposed an approach to detect and collect payloads of unknown and self-replicated polymorphic worms in homogeneous and heterogeneous cloud computing platforms. The proposed approach attaches honeypots with the inspected virtual networks to get the payload of the running worms. Different polymorphic worm instances are collected using double honeypot. Honeypots' VMs are inspected in hypervisor layer using open-source forensic tools together with virtual disk image management tools. Finally, we evaluated the proposed approach based on the previously defined challenges.

This paper is organized as follows. Section 2 gives some background of computer worms. Section 3 gives a short overview of related works. Section 4 presents our approach and illustrates the design and implementation of the proposed approach. The proposed approach is evaluated in section 5. Section 6 concludes the paper.

## II. COMPUTER WORMS

A worm is a self-replicated malware computer program, which propagated over the network with or without human intervention. Worms send copies of their code to victim machines and gain control of the execution of remote programs to execute their code in remote machines. Worms gain control of a program's execution by injecting new code into the program, injecting new control-flow edges into the program, or corrupting data used by the program [14].

The most popular techniques to detect computer worms can be categorized into two major categories [15], [7]: anomaly-based detection and signature-based detection. Anomaly-based detection techniques use the knowledge of what is considered as a normal behavior to identify computer worms. Although, anomaly-based detection is able to detect new and unknown computer worms, it reduces the network speed by checking every packet and reduces the computer performance by monitoring all activities.

Signature-based detection techniques detect known computer worms from other programs by comparing them against signatures of known computer worms. A signature is a pattern generated using instances of known computer worm to identify it. However, signature-based detection cannot detect new computer worms and maintaining signature database is not easy task [15].

Detection of computer worms can be improved by combining techniques from both categories. For example, a signature-based detection technique can be used to filter know computer worms and an anomaly-based detection technique can be used to detect unknown computer worms and collect their payloads. Collected payloads are forwarded to a signature generator, which generates signatures. Based on the generated signatures, signature-based detector will detect these worms in the future. By inspecting the behavior of the collected payloads, we can gain sufficient knowledge about their malicious activities, which will help in decontaminating infected VMs.

Worms use different techniques to be hidden and undetectable. For Example, to avoid detection by analyzing network traffic patterns, Worms blend their attack traffic within normal traffic. To avoid detection by monitoring the running processes and the loaded modules, worms attach themselves with trusted processes and modules. To avoid detection by monitoring the scanned IP addresses, worms collect addresses in infected computers to find new victims instead of scanning complete IP address space. To circumvent detection by monitoring changes in file system, worms may remain in memory for a period of time and perform some malicious activities, such as opening a back door on the host or sending the confidential information on the host to the worms' writers before modifying the file system [19]. To evade detection by memory scanners that scan the inspected memory linearly, worms move their memory structures to a region that has already been scanned [19].

To evade detection by detectors, worms try to identify if the remote system is a detector before sending their code and try to infer if they are running inside detection environment before starting their malicious actions. Worms can identify detectors remotely by analyzing its responses to network messages and identify detectors locally by monitoring their behavior.

To avoid detection by anomaly-based detectors, worms can try to be smooth and silent by performing small amounts of malicious activities and avoid loud activities (e.g., consuming the computer resources, sending a large number of Emails, establishing a large number of network connections, etc.).

To avoid detection by signature-based detectors, worms use polymorphism. Polymorphic worms change their payloads in every infection using encryption and code obfuscation. This change makes it difficult to generate signatures for these worms. Generating signature for a polymorphic worm requires collecting a large number of payload's instances and identifying the common byte strings in these payloads to be used as a worm signature. However, polymorphic worms can change their payload completely, which make it impossible to generate signatures [14]. Furthermore, polymorphic worms can avoid payloads' collectors (e.g., double-honeynet systems) by checking if the worm is already running on the remote system before sending their code.

Another technique to evade detection by signature-based detectors is designing worms as a set of components. Each standalone component is not harmful by itself and can stay in your computer without any malicious actions. Different combinations of these components perform Worm's activities. Worms are designed with a large number of combinations to make it difficult to collect their pattern [8].

Finally, we can identify the following challenges for detecting and collecting worms' payloads in cloud computing:

- Polymorphic worms: the proposed approaches have to be able to collect a large number of payload instances for the same worm even if the worm uses polymorphism technique or checks its existence in the target machine before sending its code [15].
- Customer privacy: the proposed approaches have to avoid violating the customers' privacy, which can be violated by accessing the VM memory, VM disk image, or running VM processes.
- Computational resources: the proposed approaches have to provide effective technique to detect and collect worms with minimal computational resources [16].
- Worms spreading: the proposed approaches have to avoid usage of tools that can be attacked by worms and accelerate worms spreading.
- Out-box inspection: the proposed approaches have to use out-box inspection tools (e.g., virtual machine introspection (VMI), Virtual machine monitor (VMM), etc.) instead of using in-box inspection tools that inspect VM from inside to avoid detecting or attacking them by worms [4].

- Homogeneous and heterogeneous platforms: cloud providers provide different platforms to gain a large number of customers. Customers customize the provided virtual servers by installing specific software to achieve their businesses requirements. The proposed approaches have to be able to detect and collect worms in a wide range of platforms, which is existed due to the big variation between customers' requirements. Different platforms means different types of worms because worms' writers usually write their worms based on specific vulnerabilities.
- Cloud computing management techniques: The proposed approaches have to be able to work with different cloud computing management techniques such as VM migration. In cloud computing, VM migration is used to consolidate VMs to the minimal number of physical servers according to their current resource requirements [2]. The proposed approach has to inspect VMs even if their locations are dynamically changed.

Scalability: Cloud computing datacenters are large-scale datacenters. The proposed approaches have to be scalable to work with such centers.

### III. RELATED WORK

In recent years, we have observed a rapid growth of using VMI in intrusion detection instead of using the legacy tools. In traditional approaches [10], [14], [15], [17], inspection tools are installed and running inside VMs, which are vulnerable to attack from attackers and malware programs. Over the last years, many approaches have been proposed to inspect VM from outside to improve inspection and avoid attaching [3], [5], [11], [18]. In this section, we are going to overview some of previous approaches.

The authors in [4] proposed an anomaly-based intrusion detection mechanism, which exploited virtual machine introspection (VMI) to collect a list of the running processes or modules of the running virtual machines. All collected information are gathered and analyzed by a centralized anomaly detector on a single virtual machine. Any unknown process or module that occurs on an increasing number of virtual machines is considered as a computer worm. The authors proposed a technique to collect information about unknown harmless processes. Although, using single point to gather information provides an abstract centralized view on entire cloud network, it congests network links because cloud network maybe contains hundreds of thousands of VMs. Furthermore, the proposed approach cannot detect computer worm that attached itself to trusted process.

The authors in [21] proposed a distributed network of detection agents to coordinate detection across whole cloud network. Each agent is deployed in virtual machine monitor (VMM) of a physical cloud server. Using VMI, agents collect information of the running VMs on servers they are associated with. Agents monitor VM disk image, network activity, and state of VM memory. Agents share information to determine threats that crosscut physical servers. The proposed approach reduces the load on the network links by distributing the detection. However, monitoring all running VMs in the cloud data centers is a resources consuming process and increases the cost of cooling and operating these data centers, which will decrease the income profit.

The authors in [8] identified a set of symptoms to indicate the possibility of malicious activities. For each symptom, a Forensic virtual machine (FVM) is defined to look for this symptom in running VMs. FVMs use VMI to inspect running VMs from outside. If a symptom is discovered in any VM, other FVMS are directed to inspect this VM for other symptoms. Collected information is sent to a centralized module to analyze them and take further actions.

In [19], the authors proposed high-interaction honeypot monitor, which inspects memory of the running virtual machine on Xen by using the open source forensics tool Volatility. *Volatility* accesses the VM's memory by using interface functions provided by *LibVMI API*. *LibVMI* provides interfaces for low-level functionalities provided by Xen hypervisor. Changed binaries in the inspected VM have been extracted by using *Libguestfs*, which is an open source library to analyze and manipulate file system. Although the proposed approach succeeded in capturing a wide range of malware including normal worms, it did not provide a technique to collect polymorphic worms.

In [14], the authors proposed a high-interaction double honeynet to collect polymorphic worms and proposed algorithms to generate signatures for the collected worms. However, the authors used in-box introspection tools (tools installed in inspected VMs) such as *Honeywall, Sebek*, and *Snort-inline*, which are susceptible to be detected or attacked by worms. Furthermore, the proposed approach cannot collect polymorphic worms that check their existence in the target host before sending their code because the proposed double honeynet works by forming a loop to collect all polymorphic worm instances.

In [16], the authors started by using VMI to detect incoming files and objects, and trigger file clustering module, which classifies incoming files and monitors their execution. Two algorithms have been proposed to detect malicious activities for incoming files and objects. Using VMI, data and information flow of malicious processes have been recorded instead of recording all data or information flow of all processes to simplify the analysis task. However, the proposed approach inspects all incoming files and object, and monitors their execution to detect malicious activities, which consumes many computer resources.

As was shown previously, most of current approaches either inspect VMs from inside, inspect all running VMs, inspect all running processes and modules, or cannot collect polymorphic worm instances. In the next section, we will introduce our approach to collect polymorphic worms.

IV. THE PROPOSED APPROACH DESIGN AND IMPLEMENTATION

The aim of the proposed approach is detecting and collecting payloads of zero-day active polymorphic worms in homogeneous and heterogeneous cloud computing platforms without violating the customer privacy.

In the proposed approach, worms' spreading is limited by isolating traffic of each virtual network from the traffic of other networks by using virtual switches in *XenServer* [6] or in *VMware vCenter Server* [20]. For example, Fig. 1 shows the isolation between two virtual networks using virtual distributed switches in *VMware vCenter Server*. Each virtual network contains two virtual servers, which belong to the same distributed port group. A distributed virtual switch has been created with one distributed port group for each virtual network. Virtual servers are located in ESXi hosts. Although, VM1 and VM3 are located at the same physical server and there traffics are transmitted across the same underlying NIC, they cannot sniff or inject traffic into each other because each of them belongs to different distributed port group and different distributed switch.

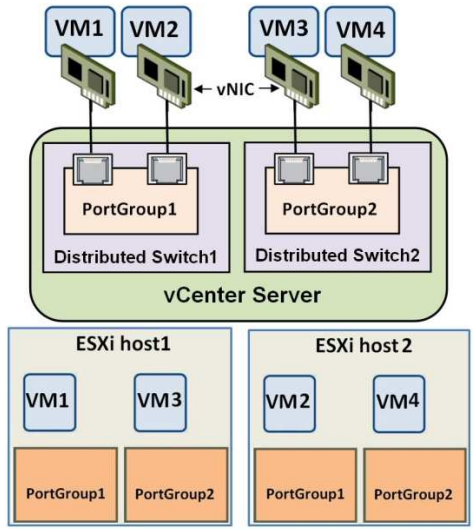

Fig. 1 Isolation between virtual networks

Fig. 2 shows the proposed approach architecture to detect and collect polymorphic worms in such networks. Introspection controller creates virtual machine inspectors from virtual machine inspectors' templates. Introspection controller assigns a set of virtual networks to each VM inspector. Introspection controller identifies current virtual networks by retrieving a list of virtual distributed switch from *vCenter server database*.

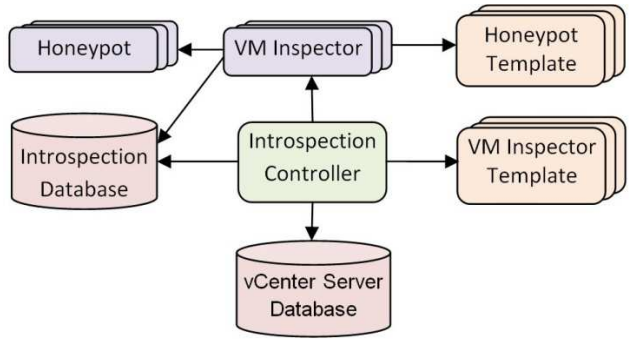

Fig. 2 The proposed approach architecture

VM inspector creates honeypot VMs from honeypot templates, which are VM templates. Each honeypot template contains a subset of software from the most common software deployed in cloud computing network. This partitioning minimizes the size of honeypots and allows each honeypot to capture a specific type of worms that are related to this subset of software. VM inspector connects each honeypot with one virtual network by adding honeypot to the port group of that virtual network (see Fig. 3). VM inspector creates and connects honeypots sequentially to avoid consuming a large number of computer resources.

If the virtual network is infected by active worms, the worms will discover the honeypot and contaminate it. After a considerable amount of time, VM inspector disconnects honeypot from the virtual network and connects it with another instance of honeypot, which is created from the same honeypot template (see Fig. 4). These two honeypots form a double honeynet. This technique reduces false positive and false negative because only active worms can discover and infect the new honeypot.

Honeypot template is designed to snapshot the running VM periodically and automatically. VM inspector analyzes the memory snapshots of the running honeypots using *Volatility* [22], which is one of the famous forensic tools. Using *Volatility*, VM inspector can gain a lot of information regarding the running VMs. For example, VM inspector can list the running processes, the previously terminated processes, hidden or unlinked processes, and the loaded DLLs for the running processes. All these lists are saved in a separated text file and used as input to the *Checker* component. The *Checker* compares these lists with previously saved lists collected from normal running to detect anomaly processes or modules.

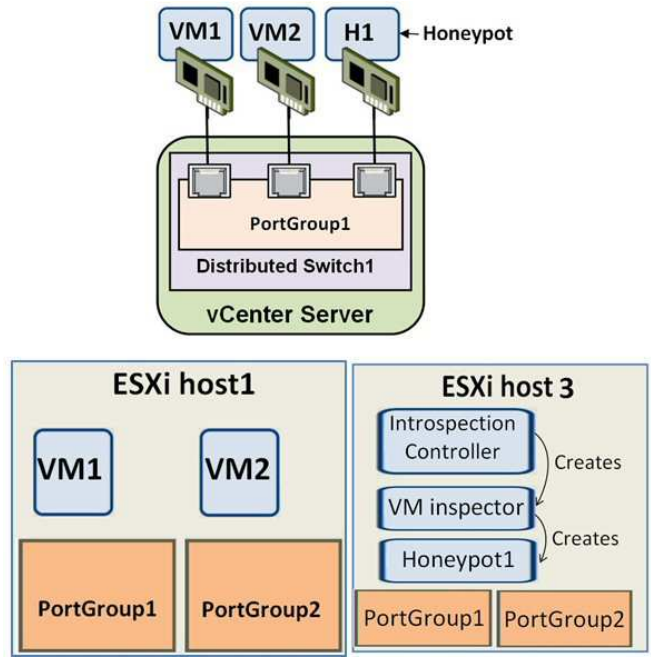

Fig. 3 Honeypot1 was created and connected with virtual network

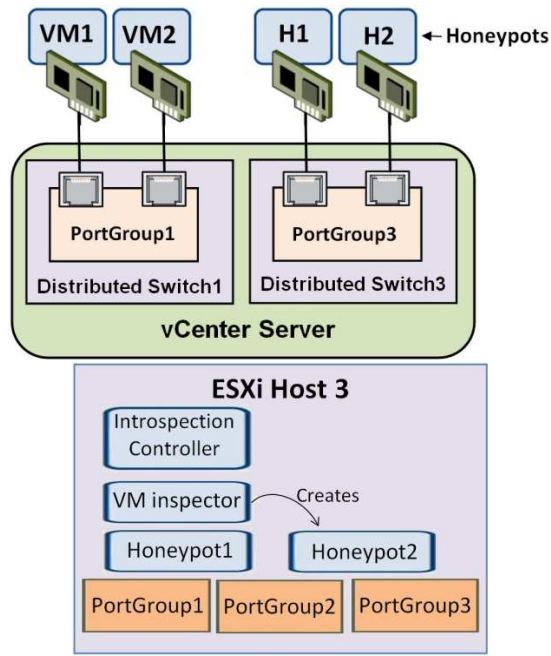

Fig. 4 Honeypot2 was created and connected with Honeypot1

Honeypot VM template is designed with independent and non-persistent hard disk. This feature in Vmware allows running VM to save any modified blocks to a separated log file (.REDO files). VM inspector uses *Virtual Disk API*, or *VixDiskLib* functions (such as *QueryChangedDiskAreas*) with *log* and *VMDK* files to retrieve the list of disk sectors that have been changed and to retrieve the list of files that have been modified, created, added, or deleted.

Using snapshots, VM inspector restores honeypots disk data and virtual machine state to the original data and state in rotate way. This restoration allows worms running in honeypots to re-contaminate honepots again if they check their existence in the target machine before sending their code. VM inspector stores all collected data in the introspection database to be used in the future by signature generators. After a considerable number of iterations, VM inspector stops the loop, destroys honeypots, and creates another honeypot to inspect the next virtual network.

## V. EVALUATION

Now, we return to the previously listed challenges in section 2 to show that the proposed approach overcome these challenges.

- Polymorphic worms: the proposed approach collects polymorphic worms payload using double honeynet, and periodically restores honeypots to the original state (before contamination) to collect worms that check if they are running in the target machine before sending their code.
- Customer privacy: to avoid violating the customers' privacy, the proposed approach does not inspect the customers' VMs directly. The proposed approach uses honeypots to allow existing worms (if any) to infect them and inspect the infected honeypots instead of inspecting the customers' VMs.
- Worms spreading: honeypots are created from clean VMs' templates and each honeypot is connected with only one virtual network to avoid transferring worms between virtual networks. All inspection tools are running outside the inspected honeypots (in hypervisor layer). Therefore, worms cannot detect or contaminate them.
- Out-box inspection: in the proposed approach, VMs inspectors inspect honeypots from the outside by inspecting snapshots. Therefore, worms cannot manipulate or detect them.
- Homogeneous and heterogeneous platforms: the proposed approach uses different VMs' template for honeypots to achieve variations between virtual networks. Each VM template is designed to capture a category of worms that are designed to use vulnerabilities in its software.
- Cloud computing management techniques: in the proposed approach, VMs inspectors connect honeypots with virtual networks by adding them to the port groups in virtual distributed switches, which allows honeypots to keep on connected with virtual networks even if they are migrated between hosts.
- Scalability: introspection controller creates and destroys VM inspectors based on the number of the running virtual networks to follow the growing and the shrinking of the cloud computing datacenters.
- Computational resources: the proposed approach does not inspect all running virtual machines. Each VM inspector uses one honeypot for each virtual network and inspects virtual networks sequentially to minimize computer resources used during inspection process. To accelerate the introspection, all created VM inspectors running concurrently. The proposed approach reduced the time required to analyze file system by analyzing log and REDO files that contain only modified sectors instead of analyzing whole VM disk image. For Example, in our implementation, we created four honypot VM templates with the specifications shown in table 1. Fig. 5 shows the size of REDO (log) files created during running VMs from these templates. As shown in Fig. 5, the size of REDO file is not comparable with the size of the VM disk image and we can reduce its size by reducing the connection time before restoring the original data. For example, in honypot VM template for *Windows Xp professional*, we analyze the log files (around 40 megabyte) instead of analyzing VM disk image with 5 GB.

TABLE I

TEMPLATES' SPECIFICATIONS FOR THE CREATED HONEYPOT VMS

| Operating System | RAM | Hard Disk | processors |
|---|---|---|---|
| Windows Xp professional | 512 MB | 5 GB | 1 |
| Windows server 2003 Enterprise Edition | 1GB | 20 GB | 1 |
| Red hat Linux | 1GB | 30 GB | 1 |
| Windows server 2008 | 1GB | 30 GB | 1 |

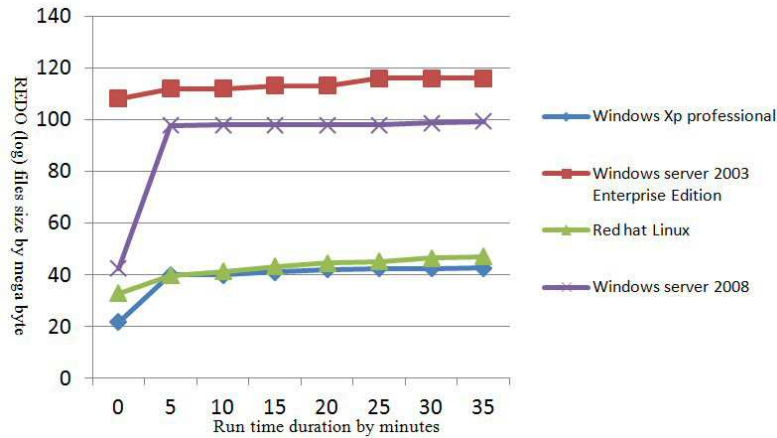

Fig. 5 The size of log files that are created by the running VMs

## VI. Conclusion

In this paper, we have identified challenges facing current approaches during collecting polymorphic computer worms in cloud computing. A high-interactive double honeypot has been proposed to address the identified challenges. The proposed approach inspects VMs from outside to detect hidden processes and to avoid detection by worms. Customers' privacy has been conserved by avoiding directly inspecting their VMs. Used computer resources have been reduced by inspecting cloud datacenters by virtual network instead of by VM. The proposed approach keeps tracking VMs even if there locations are changed over time.

In our future work, we plan to investigate how to reduce the time of inspecting VM memory and how to reduce the number of honeypots required to inspect virtual networks. Furthermore, we plan to extend our approach to inspect network traffic beside memory and file system.